\documentclass[twocolumn,aps,showpacs,prc]{revtex4}
\usepackage[dvips]{epsfig}

\begin{document}	

\title{Dilepton to photon ratio, a viscometer of  QGP}
 
\author{A. K. Chaudhuri}
\email[E-mail:]{akc@veccal.ernet.in}
\affiliation{Variable Energy Cyclotron Centre,\\ 1/AF, Bidhan Nagar,
Kolkata 700~064, India}
\author{Bikash Sinha}
\email[E-mail:]{bikash@veccal.ernet.in}
\affiliation{Variable Energy Cyclotron Centre,\\ 1/AF, Bidhan Nagar,
Kolkata 700~064, India}

\begin{abstract}
In the Israel-Stewart's 2nd order hydrodynamics, a viscous effect on   dilepton emission from a QGP medium is investigated. Dileptons are strongly affected by QGP viscosity. Large invariant mass dileptons, due to their lower velocity, are less  affected by viscosity than the low invariant mass dileptons.   We also show that the ratio of photon to dilepton is sensitive to the viscosity and can serve as a viscometer for QGP. 
 \end{abstract} 

\pacs{ PACS numbers(s):25.75.-q,12.38.Mh} 

\maketitle

\section{Introduction}

It is expected that collisions between two nuclei at ultra-relativistic energies will lead to a phase transition from hadrons to the fundamental constituents, quarks and gluons, usually referred to as Quark-Gluon-Plasma (QGP). Although the nature of the phase transition still remains largely uncertain, it is an acceptable reality that QGP is formed.

Electromagnetic signals, photons and dileptons are excellent probe of QGP, formed immediately after the collision of the two nuclei. The interaction, being primarily electromagnetic, photons and dileptons tend to retain the information of early times rather more efficiently compared to hadronic signals; the hadrons interact strongly and thus tend to erase the information of early time retaining the information only around the freeze-out surface.


In ideal hydrodynamic models, photon and dilepton production in high energy nuclear   energy has been studied extensively \cite{Alam:1992gw,Sarkar:2002ui, Alam:2007dv,Chatterjee:2005de}. However, it is now realized that the strongly interacting medium, produced in Au+Au collisions, must be treated as a viscous medium. Gravity dual (AdS/CFT) theories suggest that specific viscosity, i.e. viscosity to entropy ratio of {\em any matter} has a lower bound, the so called KSS bound $\eta/s=1/4\pi$ \cite{Policastro:2001yc,Kovtun:2003wp}. Even though, photons and dileptons are important probe of QGP matter,
 viscous effects on these signals still need to be investigated extensively.   Dusling  \cite{Dusling:2008xj,Dusling:2009bc} studied viscous effects on these signals. In a recent paper \cite{Chaudhuri:2011up}, we have investigated viscous effects on photons produced from the QGP medium. It was shown that viscous effects on photons are strong. 
It was suggested that the slope parameter of the transverse momentum spectra for photons can possibly limit the thermalisation time and viscosity. In the present paper, we investigate viscous effect on dilepton emission from QGP. Unlike photons, dileptons are quite massive. Invariant mass of the dileptons make them unique probe of QGP. In addition to transverse momentum, the invariant mass of the dileptons can be tuned to explore very early QGP. For example, dileptons of invariant mass $M_{inv}\sim$ 1 GeV are expected to be formed in the time scale $\tau_{for} \sim 1/M_{inv}\sim$ 0.2 fm and  can explore the QGP at very early time $\tau\sim$ 0.2 fm. Viscous effects on dileptons will also depend on the invariant mass. Viscous drag on dileptons is proportional to relative velocity between the fluid and the lepton pairs. Depending on the invariant mass, viscous drag will vary. As will be shown later, this unique feature makes them very sensitive to the viscosity. Interestingly enough, we find in this work that the ratio, dileptons to  photons is quite sensitive to the viscosity and thus the ratio turns out to be an excellent viscometer.

\section{Hydrodynamical equations, equation of state and initial conditions}
 
In a hydrodynamical model, the invariant distribution of dileptons  is obtained by convoluting the  dileptons production rate with space-time evolution of the fluid. We assume that in $\sqrt{s}_{NN}$=200 GeV, Au+Au collisions  at RHIC, a baryon free QGP fluid is formed. Space-time evolution of the fluid is obtained by solving 2nd order Israel-Stewart's theory,

\begin{eqnarray}  
\partial_\mu T^{\mu\nu} & = & 0,  \label{eq1} \\
D\pi^{\mu\nu} & = & -\frac{1}{\tau_\pi} (\pi^{\mu\nu}-2\eta \nabla^{<\mu} u^{\nu>}) \nonumber \\
&-&[u^\mu\pi^{\nu\lambda}+u^\nu\pi^{\nu\lambda}]Du_\lambda. \label{eq2}
\end{eqnarray}

Eq.\ref{eq1} is the conservation equation for the energy-momentum tensor, $T^{\mu\nu}=(\varepsilon+p)u^\mu u^\nu - pg^{\mu\nu}+\pi^{\mu\nu}$, 
$\varepsilon$, $p$ and $u$ being the energy density, pressure and fluid velocity respectively. $\pi^{\mu\nu}$ is the shear stress tensor (we are neglecting bulk viscosity). Eq.\ref{eq2} is the relaxation equation for the shear stress tensor $\pi^{\mu\nu}$.   
In Eq.\ref{eq2}, $D=u^\mu \partial_\mu$ is the convective time derivative, $\nabla^{<\mu} u^{\nu>}= \frac{1}{2}(\nabla^\mu u^\nu + \nabla^\nu u^\mu)-\frac{1}{3}  
(\partial . u) (g^{\mu\nu}-u^\mu u^\nu)$ is a symmetric traceless tensor, $\eta$ is the shear viscosity and $\tau_\pi$ is the relaxation time.  It may be mentioned that in a conformally symmetric fluid relaxation equation can contain additional terms  \cite{Song:2008si}.
Assuming boost-invariance, Eqs.\ref{eq1} and \ref{eq2}  are solved in $(\tau=\sqrt{t^2-z^2},x,y,\eta_s=\frac{1}{2}\ln\frac{t+z}{t-z})$ coordinates, with the code   "`AZHYDRO-KOLKATA"', developed at the Cyclotron Centre, Kolkata.
 Details of the code can be found in \cite{Chaudhuri:2008sj,Chaudhuri:2009uk,Chaudhuri:2008ed}. 

 Eqs.\ref{eq1},\ref{eq2} are closed with an equation of state $p=p(\varepsilon)$.
Lattice simulations      \cite{Cheng:2007jq,Cheng:2009zi,Aoki:2006br,Aoki:2009sc}
indicate that the confinement-deconfinement transition is a cross over, rather than a 1st or 2nd order phase transition. There is no critical temperature for a cross-over transition.  
The inflection point on the Polyakov loop is generally defined as the pseudo critical temperature. In Wuppertal-Budapest simulation \cite{Aoki:2006br,Aoki:2009sc}, the pseudo critical temperature 
$T_c\approx$170 MeV. In the present simulation, for the QGP phase, we use an EOS based on Wuppertal-Budapest simulation.

Solution of partial differential equations (Eqs.\ref{eq1},\ref{eq2}) requires initial conditions, e.g.  transverse profile of the energy density ($\varepsilon(x,y)$), fluid velocity ($v_x(x,y),v_y(x,y)$) and shear stress tensor ($\pi^{\mu\nu}(x,y)$) at the initial time $\tau_i$. One also need to specify the viscosity ($\eta$) and the relaxation time ($\tau_\pi$). A freeze-out
temperature is also needed. In the following, we will study viscous effects on dilepton production from the QGP phase only. The hydrodynamical equations are then solved till the freeze-out temperature $T_F=T_c$=170 MeV. At the initial time $\tau_i$, initial energy density is assumed to be distributed as \cite{QGP3}

\begin{equation} \label{eq3}
\varepsilon({\bf b},x,y)=\varepsilon_i[(1-f_{hard}) N_{part}({\bf b},x,y) +f_{hard} N_{coll}({\bf b},x,y)],
\end{equation}

\noindent
where b is the impact parameter of the collision. $N_{part}$ and $N_{coll}$ are the transverse profile of the average participant and collision number respectively, calculated in a Glauber model. $f_{hard}$=0.13 is the hard scattering fraction \cite{Hirano:2009ah}. 
$\varepsilon_i$ is the central energy density of the fluid in impact parameter $b=0$ collision. In the following, we assume that the fluid is thermalised at the time scale $\tau_i$=0.6 fm to central temperature   $T_0$=350 MeV.
A large number of
charged particle's data e.g. identified particles spectra, elliptic flow etc.  data are explained in hydrodynamical model with similar initial time and temperature scale \cite{QGP3}. We also assume that initial fluid velocity is zero, $v_x(x,y)=v_y(x,y)=0$. The shear stress tensor was initialized with boost-invariant value, $\pi^{xx}=\pi^{yy}=2\eta/3\tau_i$, $\pi^{xy}$=0. For the relaxation time, we use the Boltzmann estimate $\tau_\pi=3\eta/4p$. We further assume that shear viscosity to entropy ratio is a constant throughout the evolution and  we simulate Au+Au collisions for a range of $\eta/s$, $\eta/s$=0-0.12.

\section{Dilepton production rates}

The rate of dilepton emission, from a QGP, due to quark-antiquark annihilation can be written as, 

\begin{widetext}
\begin{equation}
\frac{dN}{d^4q}=\int \frac{d^3p_1}{(2\pi)^3} \frac{d^3p_2}{(2\pi)^3} 
 f_1(E_1,T)f_2(E_2,T) v_{12} \sigma(M^2)
  \delta^4(q-p_1-p_2)  \label{eq4}
\end{equation}
\end{widetext}

\noindent where $q=(q_0,\bf{q})$ is the virtual photon 4-momenta and $M^2=(E_1^2+E_2^2)-(\bf{p_1}^2+\bf{p_2}^2)$ is the invariant mass. In Eq.\ref{eq4}, $v_{12}=\frac{M^2}{2E_1 E_2}$ is the relative velocity between the quark-antiquark pair and $\sigma=\frac{16\pi\alpha^2 e_q^2 N_c}{3M^2}$ is the $q\bar{q}$ cross-section. $f(E,T)$ in Eq.\ref{eq4} is the quark/antiquark distribution function.  
In ideal fluid evolution, QGP is in (local) equilibrium and  $f(E,T)=f_{eq}(E,T)=\frac{1}{1+e^{E/T}}$. 
For $E/T=p.u/T << 1$ the distribution function can be approximated by a Boltzmann distribution and
integrations in Eq.\ref{eq4} can be analytically performed, 

\begin{equation}
\left (\frac{dN}{d^4q} \right )_{eq}=\frac{N_c\alpha^2 e_q^2}{12\pi^4} e^{-q_0/T}
\end{equation}

  \begin{figure}[t]
\center
\resizebox{0.35\textwidth}{!}{%
  \includegraphics{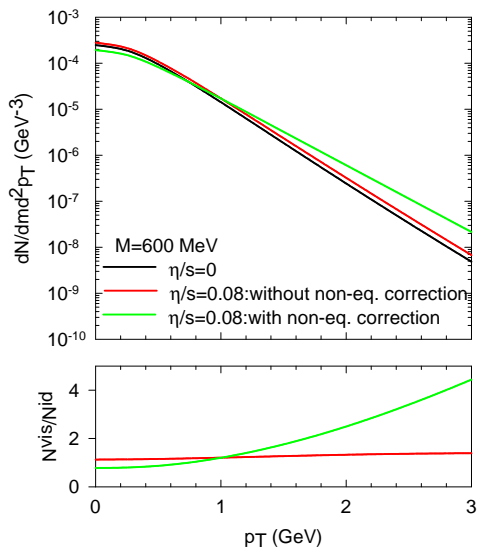}
}
\caption{(color online)
Transverse momentum spectra for dileptons of invariant mass $M$=600 MeV, from evolution of ideal and minimally viscous ($\eta/s$=0.08) QGP. For viscous fluid, spectra obtained with and without the  non-equilibrium correction to distribution function, are shown separately. In the bottom panel, ratio of production from viscous fluid with and without the non-equilibrium correction to the ideal fluid is shown.
 }\label{F1}
\end{figure}

Unlike an ideal fluid,   
in viscous evolution,   distribution functions $f$ is modified due to non-equilibrium correction. For small non-equilibrium effect, the distribution function can be approximated as,

\begin{equation}
f_{neq} \rightarrow f_{eq}(1+\delta f_{neq}), \delta f_{neq} << 1.
\end{equation}

The non-equilibrium correction $\delta f_{neq}$ depend on dissipative forces as well as on particle momenta. For shear viscosity, non-equilibrium correction to equilibrium distribution can be obtained from the following ansatz,

\begin{equation} \label{eq6}
\delta f_{neq}=C q_\mu q_\nu \pi^{\mu\nu}=\frac{1}{2(\varepsilon+p)T^2}q_\mu q_\nu \pi^{\mu\nu}
\end{equation}

Dilepton production rate from non-equilibrium QGP then has two parts, an equilibrium part and a non-equilibrium part, 

\begin{equation}
\left (\frac{dN}{d^4q}\right )= \left (\frac{dN}{d^4q}\right )_{eq} 
\left (1+\left (\frac{d\delta N}{d^4q}\right )_{neq} \right )
\end{equation}

Approximating the equilibrium distribution function by Boltzmann distribution,
Dusling and Lin \cite{Dusling:2008xj} evaluated the non-equilibrium correction to the dilepton production rate, 

\begin{equation}
\left (\frac{d\delta N}{d^4q}\right )_{neq}=   \frac{q_\mu q_\nu \pi^{\mu\nu}}{3(e+p)T^2}
\end{equation}

  Since non-equilibrium correction to distribution function is assumed to be small, it is essential that,

\begin{equation}
\left (\frac{d\delta N}{d^4q} \right )_{neq} <<  1.
\end{equation}

\begin{figure}[t]
\center
\resizebox{0.35\textwidth}{!}{%
  \includegraphics{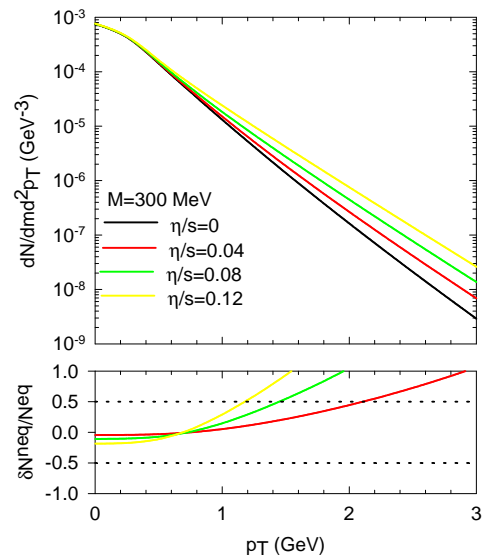}
}
\caption{(color online) Transverse momentum spectra for dileptons of invariant mass M=300 MeV, form evolution of QGP fluid with viscosity to entropy ratio $\eta/s$=0, 0.04, 0.08 and 0.12. In the lower panel, ratio of non-equilibrium correction to equilibrium production is shown }\label{F2}
\end{figure}     

\section{Effect of viscosity on dilepton spectra}

For viscosity to entropy ratio $\eta/s$=0, 0.04, 0.08 and 0.12, we have simulated 20-40\% Au+Au collisions and computed dileptons invariant distribution.  In Fig.\ref{F1}, transverse momentum spectra for dileptons of invariant mass $M$=600 MeV,   from evolution of ideal and minimally viscous ($\eta/s$=0.08) QGP fluid, are shown.  In viscous evolution, dilepton spectra is modified due to (i) changed
space-time evolution and (ii) non-equilibrium correction to the equilibrium distribution function.  
To understand the effect of viscosity, spectra, with or without the  non-equilibrium correction to the equilibrium distribution function are shown separately. In the lower panel of Fig.\ref{F1},  the ratio dileptons from viscous QGP  and ideal QGP are shown.  If the non-equilibrium correction to the equilibrium distribution function is neglected, in viscous evolution dilepton yield increase by a factor of $\sim$1.3 . The increase is largely $p_T$ independent. The ratio become $p_T$ dependent  when the non-equilibrium correction is included. It is expected that the non-equilibrium correction increases   quadratically with $p_T$ (see Eq.\ref{eq6}).

Simulated transverse momentum spectra for dileptons of invariant mass M=300 MeV,  for four values of viscosity to entropy ratio, $\eta/s$= 0 (ideal fluid), 0.04, 0.08 (AdS/CFT lower bound) and 0.12 are shown in Fig.\ref{F2}. 
In viscous evolution, dilepton spectra is stiffened, stiffening is more for more viscous fluid. The result is understood. In viscous evolution, non-equilibrium correction introduces additional momentum dependence which is linearly proportional to shear stress tensors. 
One also note that at very low $p_T$, yield is marginally reduced in viscous fluid. This result is not unphysical.  At low $p_T$ non-equilibrium correction to the equilibrium distribution function can be negative. A similar decrease is seen in case of low $p_T$ hadron production also \cite{Teaney:2003kp}. 
 
As it was mentioned earlier, it is essential that the non-equilibrium correction to dilepton production is small compared to the equilibrium production. The non-equilibrium distribution function is defined under the assumption that the system is not far away from equilibrium and the non-equilibrium correction to the equilibrium distribution function is small ($f_{neq} <<1$).   In the lower panel of Fig.\ref{F2}, the ratio of non-equilibrium correction to dilepton production to the equilibrium production is shown. At very low $p_T$, $p_T <0.6 MeV$, non-equilibrium correction is negative but small. At larger $p_T$ correction is positive and rapidly increase with $p_T$. For viscosity to entropy ratio $\eta/s$=0.04, 0.08 and 0.12, non-equilibrium correction equals  the equilibrium production at $p_T\approx$ 2.9, 1.9 and 1.5 GeV respectively. Definitely, viscous hydrodynamics is inapplicable beyond those $p_T$ range.
Indeed, if we assume that hydrodynamics remain applicable till $|\delta N^{neq}/N^{eq}| \leq 0.5$, the dilepton production from   QGP, with viscosity to entropy ratio, 0.04, 0.08 and 0.12 can be computed with confidence only up to $p_T$=(0-2.1 GeV), (0-1.5 GeV) and (0-1.2 GeV). 
The result is similar to that obtained in our simulations for photon production in viscous hydrodynamics \cite{Chaudhuri:2011up}. Viscous effects are also very strong on photon and photon production can be computed only in a limited $p_T$ range.

 \begin{figure}[t]
\center
\resizebox{0.35\textwidth}{!}{%
  \includegraphics{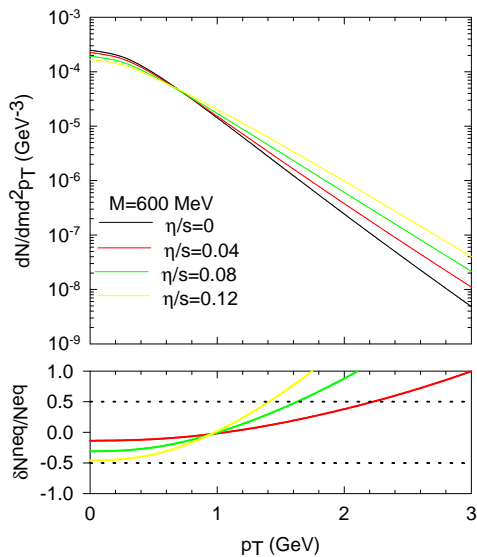}
}
\caption{(color online) same as in Fig.\ref{F2} but for dilepton invariant mass M=600 MeV.}\label{F3}
\end{figure}   

In Fig.\ref{F3}   same results are shown for dilepton invariant mass $M$=600   MeV. As before, at low $p_T$, non-equilibrium correction contribute negatively, and dilepton production is reduced in viscous evolution. The correction is positive at higher $p_T$ and dilepton yield is increased. In the lower panel, the ratio of non-equilibrium correction to equilibrium production is shown. As expected, correction increases with $p_T$ and also with viscosity. For   viscosity to entropy ratio $\eta/s$=0.04, 0.08 and 0.12, non-equilibrium corrections become equal the equilibrium contribution at $p_T\approx$ 3 GeV, 2.1 GeV and  1.7 GeV. We have not shown but
similar result is obtained for higher mass M=900 MeV dileptons.  For M=900 MeV,  
for viscosity to entropy ratio $\eta/s$=0.04, 0.08 and 0.12, non-equilibrium correction equals the equilibrium contribution at $p_T \approx$ 3.2 GeV, 2.4 GeV and 2 GeV.  It is interesting to note that the transverse momentum $p_{T_{max}}$ at which non-equilibrium correction equals to equilibrium correction, in addition to viscosity, also depend on the dilepton invariant mass.  For example, for $\eta/s$=0.12, non-equilibrium contribution equals the equilibrium contribution at  $p_{T_{max}}$= 1.5 GeV, 1.7 GeV and 2 GeV respectively for dilepton invariant mass M=300, 600 and 900 MeV. It appear that at large $p_T$ viscous effect is lessened in more massive dileptons. The result can be understood qualitatively. Dilepton velocity is approximated as $1/\sqrt{M^2+p_T^2}$. More massive dileptons will move slowly, consequently, viscous drag on massive dileptons is less than on lighter dileptons. The feature is important. It will be shown later this particular feature can be utilised to accurately measure QGP viscosity.

    \begin{figure}[t]
\center
\resizebox{0.35\textwidth}{!}{%
  \includegraphics{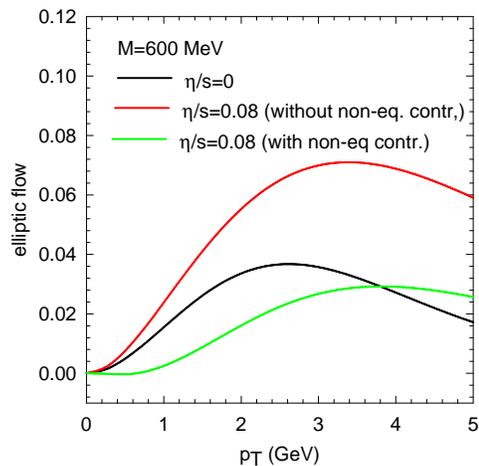}
}
\caption{(color online) elliptic flow in ideal and viscous QGP. For viscous QGP, flow with and without the non-equilibrium correction is shown separately.}\label{F4}
\end{figure}

\section{Effect of viscosity on dileptons elliptic flow}
 
Elliptic flow is an important observable in relativistic energy nuclear collisions. Large elliptic flow of observed hadrons is indirect proof of thermalisation of the system.  
In Fig.\ref{F4}, we have shown the simulation results for   elliptic flow for dileptons of invariant mass M=600 MeV. Results are shown for ideal and viscous   $\eta/s$=0.08 evolution. Flow with and without the non-equilibrium correction is shown separately to understand the effect of viscosity.
In ideal fluid evolution, elliptic flow increases with increasing $p_T$ up to $p_T\approx$2.5 GeV. At larger $p_T$, flow decreases, indicating that high $p_T$ dileptons are overwhelmingly from the early   non-flow phase. Elliptic flow dileptons is not large, even in ideal QGP, in 20-40\% collision,  the highest $v_2(p_T=2.5 GeV)\approx 0.04$. Interestingly, if the non-equilibrium correction  is neglected, in viscous fluid evolution, elliptic flow increases. However, if the non-equilibrium correction is included, the simulation correctly predicts reduced flow. The result indicates the importance to have a consistent theory. Entirely wrong conclusion can be reached in an inconsistent theory.  

\begin{figure}[t]
\center
\resizebox{0.35\textwidth}{!}{%
  \includegraphics{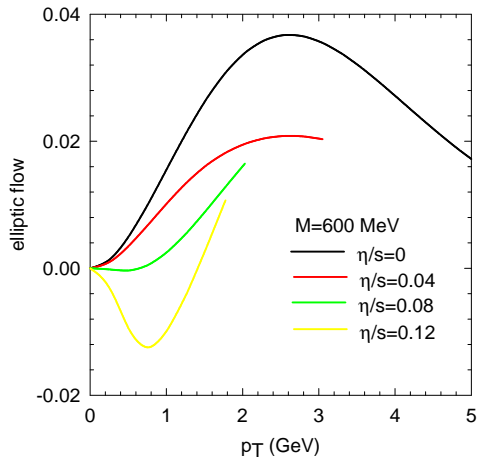}
}
\caption{(color online) Elliptic flow for dileptons of   invariant mass M=600 MeV,
 as a function of QGP viscosity.  }\label{F5}
\end{figure}

In Fig.\ref{F5}, elliptic flow dileptons of invariant mass M=600 MeV is shown as a function of viscosity. The flow is shown only in the $p_T$ range when viscous hydrodynamics is applicable. With increasing viscosity flow is reduced and become negative at large viscosity to entropy ratio $\eta/s$=0.12.

\section{Dilepton to photon ratio as viscometer for QGP}

 QGP viscosity is an important parameter. Theoretical estimates of the ratio, (shear) viscosity over the entropy density, $\eta/s$ cover a wide range,   0.0-1.0. String theory based models (ADS/CFT) give a lower bound on viscosity of any matter $\eta/s \geq 1/4\pi$ \cite{Policastro:2001yc}. In a perturbative QCD, Arnold et al  \cite{Arnold:2000dr} estimated $\eta/s\sim$ 1.  In a SU(3) gauge theory, Meyer \cite{Meyer:2007ic} gave the upper bound $\eta/s <$1.0, and his best estimate is $\eta/s$=0.134(33) at $T=1.165T_c$.  At RHIC region, Nakamura and Sakai \cite{Nakamura:2005yf} estimated the viscosity of a hot gluon gas  as $\eta/s$=0.1-0.4. Attempts have been made to estimate QGP viscosity directly from experimental data.  Gavin and Abdel-Aziz \cite{Gavin:2006xd} proposed to measure viscosity from transverse momentum fluctuations. From the existing data on Au+Au collisions, they estimated  QGP viscosity as $\eta/s$=0.08-0.30. Experimental data on elliptic flow has also been used to estimate QGP viscosity. Elliptic flow scales with eccentricity. Departure from the scaling can be understood as due to off-equilibrium effect and utilised to estimate viscosity , $\eta/s$=0.11-0.19 
\cite{Drescher:2007cd,Lacey:2006bc,Chaudhuri:2009ud,Chaudhuri:2009hj}. The estimates are well within the upper bound obtained $\eta/s < 0.5$ obtain in \cite{Luzum:2008cw,Song:2008hj}.
 However, hydrodynamical models estimates of QGP viscosity can be largely uncertain due to uncertainty in initial conditions. In a detailed analysis of RHIC data on phi meson \cite{Chaudhuri:2009uk}, it was shown that uncertainties initial conditions in hydrodynamic models can induce very large uncertainty ($\sim$175\% or more) in viscosity to entropy ratio.
 
    \begin{figure}[t]
\center
\resizebox{0.35\textwidth}{!}{%
  \includegraphics{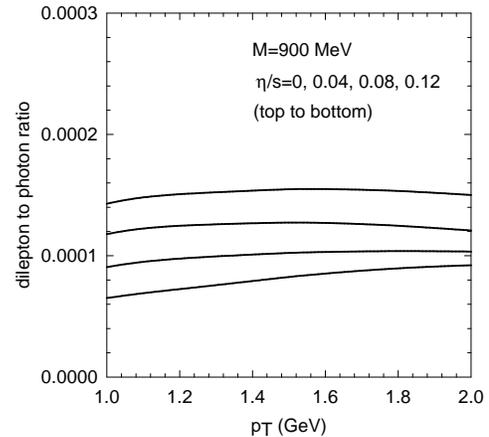}
}
\caption{For dileptons of invariant mass M=900 MeV, the ratio of dileptons to photon is shown. The four lines (top to bottom) corresponds to viscosity to entropy ratio $\eta/s$=0, 0.04, 0.08 and 0.12 respectively.}\label{F6}
\end{figure}

Ideally speaking,  hadrons are not the best probe to study QGP viscosity.  
Hadrons are emitted from the hadronic phase. They can carry viscosity information of the QGP phase only indirectly that too with the assumption that the non-perturbative hadronisation process did not erase the memory of the earlier QGP phase. Electromagnetic probes like photons and dileptons, on the other hand are emitted directly from the QGP phase and can carry directly the information of viscosity of the QGP phase. However, below the critical temperature, QGP fluid transforms into hadronic fluid.  Photons and dileptons are also produced from the hadronic phase. Experimental data then contain contributions of both the QGP and hadronic phases. Only in a limited phase space QGP phase dominates the production. For photons, QGP phase dominates the production in the $p_T$ range,   $1.5 \leq p_T \leq 2.5$ GeV  \cite{Alam:2007dv,Chatterjee:2005de}. Large invariant mass dileptons are also dominated by the QGP phase.

In a recent paper \cite{Chaudhuri:2011up}, we have studied viscous effect on direct photon production.  
Details of the calculation can be seen in \cite{Chaudhuri:2011up}. 
We now argue that the   dilepton to photon ratio can serve as an effective viscometer for QGP.  The argument is simple, spectra of both photon and dileptons are stiffened in viscous QGP. Stiffening is more in more viscous QGP. Low invariant mass dileptons behaves essentially as photons and are similarly affected by the viscosity. Viscous effects will largely cancels out and  the ratio will not be sensitive to viscosity. But large invariant mass dileptons moves slowly, viscous drag on them is less than on photons. Photon spectra will stiffen more than the large invariant mass dilepton spectra and the ratio will drop with increasing viscosity.  The ratio has the advantage that the uncertainties in initial conditions can be largely eliminated.

    \begin{figure}[t]
 \vspace{0.3cm} 
\center
\resizebox{0.35\textwidth}{!}{%
  \includegraphics{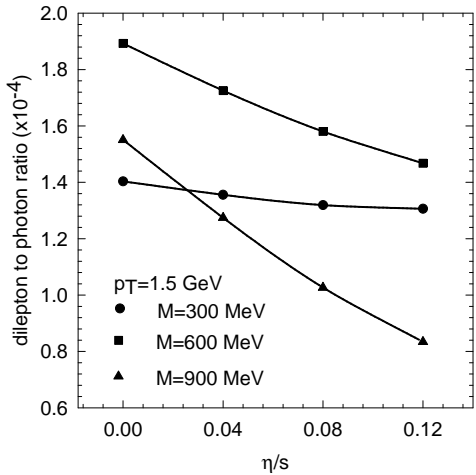}
}
\caption{For dileptons invariant mass M=300, 600 and 900 MeV, dileptons to photon ratio, at $p_T$=1.5 GeV,  is shown as a function of viscosity.  }\label{F7}
\end{figure}   

In Fig.\ref{F6}, for viscosity to entropy ratio $\eta/s$=0, 0.04, 0.08 and 0.12, the ratio of dilepton to photon production, for dileptons invariant mass $M=900 $ MeV,  is shown. We have shown the results only in the $p_T$ range in which viscous hydrodynamics remain applicable for the highest $\eta/s$=0.12 considered here. The ratio depends weakly on $p_T$, $p_T$ dependence of the non-equilibrium correction largely canceling out.  As argued before, the ratio drops with increasing viscosity. The dependence of viscosity can be seen more clearly in Fig.\ref{F7}.
In Fig.\ref{F7}, the dilepton to photon ratio at $p_T\approx$1.5 GeV is shown as a function of viscosity. For invariant mass M=300 MeV, the ratio hardly changes with of viscosity. For invariant mass M=600 MeV, the ratio decreases with viscosity, the decrease is even faster for invariant mass M=900 MeV.   Fast change in the ratio for large invariant mass dileptons 
makes it very sensitive to viscosity. For dilepton mass M=600 MeV, if the ratio is measured within 10\% accuracy, viscosity to entropy ratio can be estimated within an accuracy of   $\sim$5\%. The sensitivity is increased to $\sim$2\% for invariant mass M=900 MeV. We have shown the ratio at a fixed $p_T\approx$ 1.5 GeV. The slope of the ratio is largely $p_T$ independent and sensitivity of the ratio to viscosity remain unaltered at other $p_T$ also.  It may be mentioned here that the above discussion is based on the assumption that  $p_T\approx$1.5 GeV, photons and dielptons are dominantly from the QGP phase. While ideal hydrodynamics analysis support the assumption, it quite likely that window for QGP  may well get modified with viscous effect.  

It has been pointed out before  \cite{Sinha:1983jm,Sinha:1985vw} that the dilepton to photon ratio is a clear signal of QGP; by taking the ratio uncertainties of initial conditions and other parameters largely get cancelled out. 
Now, with the ratio being an accurate viscometer, the advantage of the ratio can be further utilised.

\section{Summary}

To summarize, we have studied viscous effects on dilepton production from QGP.
 In viscous dynamics, dilepton production is modified due to (i) changed space-time evolution of the fluid and (ii) non-equilibrium correction to the equilibrium distribution function. With initial conditions appropriate for $\sqrt{s}_{NN}$=200 GeV Au+Au collisions, space-time evolution of QGP was obtained by solving Israel-Stewart's 2nd order hydrodynamics. Invariant distribution for dileptons was obtained by convoluting the dilepton production rates over the space-time evolution. Effect of viscosity to stiffen the dilepton spectra and reduce elliptic flow. Stiffening is more in more viscous QGP. Elliptic flow is also reduced more in  more viscous fluid. It was also shown that 
viscous effect on dileptons is strong. Even for minimally viscous QGP ($\eta/s=0.08$),  dilepton production can be reliably computed, only in a limited $p_T$ range.  It is  also indicated that for large mass dileptons, the  ratio of dilepton to photon yield is sensitive to viscosity to entropy ratio. Accurate estimate of QGP viscosity can be obtained by measuring the  ratio experimentally.

\end{document}